# Optical time-harmonic elastography for multiscale stiffness mapping across the phylogenetic tree

Jakob Jordan[1], Noah Jaitner[1], Tom Meyer[1], Luca Brahmè[2,3], Mnar Ghrayeb[4,5], Julia Köppke[2,3], Stefan Klemmer Chandia[1], Vasily Zaburdaev[6,7], Liraz Chai[4,5], Heiko Tzschätzsch[8], Joaquin Mura[9], Anja I.H. Hagemann[2,3], Jürgen Braun[8], Ingolf Sack[1]

1 Department of Radiology, Charité – Universitätsmedizin Berlin
2 Department of Hematology/Oncology, Charité – Universitätsmedizin Berlin
3 German Cancer Consortium (DKTK)—German Cancer Research Center (DKFZ), 69120 Heidelberg
4 The Center for Nanoscience and Nanotechnology, The Hebrew University of Jerusalem, Edmond J. Safra Campus, Jerusalem, 91901, Israel
5 Institute of Chemistry, The Hebrew University of Jerusalem, Edmond J. Safra Campus, Jerusalem, 91901, Israel
6 Department of Biology, Friedrich-Alexander Universität Erlangen-Nürnberg, 91058 Erlangen
7 Max-Planck-Zentrum für Physik und Medizin, 91058 Erlangen, Germany
8 Institute of Medical Informatics, Charité – Universitätsmedizin Berlin
9 Department of Mechanical Engineering, Universidad Técnica Federico Santa María, Santiago, Chile.

## Abstract

Rapid mapping of the mechanical properties of soft biological tissues from light microscopy to macroscopic imaging could transform fundamental biophysical research by providing clinical biomarkers to complement *in vivo* elastography. We here introduce superfast optical time-harmonic elastography (OTHE) to remotely encode surface and subsurface shear wave fields for generating maps of tissue stiffness with unprecedented detail resolution. OTHE rigorously exploits the space-time propagation characteristics of time-harmonic waves to address current limitations of biomechanical imaging and elastography. Key solutions are presented for stimulation, encoding, and stiffness reconstruction of time-harmonic, multifrequency shear waves, all tuned to provide consistent stiffness values across resolutions from microns to millimeters. OTHE's versatility is demonstrated in *Bacillus subtilis* biofilms, zebrafish embryos, adult zebrafish, and human skeletal muscle, reflecting the diversity of the phylogenetic tree from a mechanics perspective. By zooming in on stiffness details from coarse to finer scales, OTHE advances developmental biology and offers a way to perform biomechanics-based tissue histology that consistently matches *in vivo* time-harmonic elastography in patients.

## Introduction

The shear modulus of biological tissue is a fundamental parameter that characterizes stiffness and holds the key to a plethora of insights into parenchymal biophysical structures and functions. For example, stiffness is a guiding principle for mechanosensing, which in turn enables cells to act collectively in both normal and abnormal conditions as diverse as embryonic development [1-3], tumor progression [4, 5], or microbial pattern formation [6, 7]. Mechanical cues allow cells to interact with their environment across the entire phylogenetic tree [8, 9]. In medical imaging, stiffness measurement by elastography holds great promise as a diagnostic marker in a wide range of diseases from fibrosis to cancer [10].

However, there are many challenges to stiffness measurement in biological materials, from the cellular level [11] to noninvasive elastography [12]. The need to apply external mechanical stress to elicit and measure tissue deformation and the way this stress is applied including its rate and amplitude can significantly affect the stiffness values obtained. For instance, large strain can induce compression stiffening [13], similar to the effect of higher strain rates, which increase apparent stiffness due to viscoelastic dispersion [14]. In addition, biological tissues have complex boundary conditions, making object geometry, spatial heterogeneity, substrate stiffness, and surface texture potential factors affecting measurement results [12].

To address these challenges, investigators have developed a number of methods including high-throughput single-cell deformability tests [15] and optical stretching [16], surface-based tissue methods such as atomic force microscopy (AFM) [17] and parallel-plate shear rheometry [18], nondestructive techniques such as Brillouin microscopy [19], optical coherence elastography [20-22], and shear wave-based elastography for the investigation of microsamples [23] to patients [12].

This arsenal of mechanobiological tools has complementary strengths in terms of resolution, noninvasiveness, and applicability to living systems. For example, the optical stretcher has revealed cancer cell softening in neurological tumors after tissue resection, and the results correlate with in vivo bulk tissue properties previously quantified by magnetic resonance elastography (MRE) [24]. There are other examples in the literature showing correlations between macroscopic in vivo MRE and micro-indentation-based AFM [25].

Notwithstanding these promising results, the missing link in the mechanobiology toolbox is a method that applies the same mechanical stress as *in vivo* but provides microscopic resolutions, preferably with the ability to cover larger tissue sections similar to those seen with macroscopic MRE. To fill this gap, we introduce optical time-harmonic elastography (OTHE), which combines scale-invariant optical detection of shear waves with the principles of time-harmonic elastography. A review of the long history of optical methods in elastography is provided in the supplementary material (Supplementary note 1: Optical elastography methods). Key benefits of time-harmonic elastography leveraged for OTHE include (i) cross-modality comparison of results through the use of precisely known frequencies [26], (ii) mitigated boundary-related artifacts through multifrequency direct solutions of the wave equation [12], (iii) time-efficient stroboscopic encoding [27], and motion estimation [28] optimized to sinusoidal oscillations. As a result, OTHE can map stiffness in fractions of a second with µm resolution and values that are directly comparable to well-established elastography measurements.

The versatility of OTHE is demonstrated in tissue systems as diverse as the species encountered across the phylogenetic tree: *Bacillus subtilis* biofilms, which represent archaic colonies formed by protein- and polysaccharide-secreting bacteria; zebrafish embryos, whose genes are about 80% orthologous with humans [29]; and human skeletal muscle, whose stiffness changes with activity. Biofilms are prokaryotic analogues of tissue-like structures [30] with cell differentiation, heterogeneous phenotype distribution, extracellular matrix production, and formation of intricate vasculature-like water-filled channels for nutrient distribution [31]. Among eukaryotes, zebrafish embryos provide a prototypical and versatile model for studying tissue differentiation and full organ formation [32, 33]. In vivo human skeletal muscle poses many challenges to current elastography techniques due to its multi-scale anisotropic architecture and load-dependent stiffness changes [34]. It has been argued that, for the assessment of muscle dysfunction, the degree of stiffness change of skeletal muscle during exercise is more relevant than static baseline values [35], highlighting the need for rapid methods capable of tracking dynamic changes in stiffness during exercise.

There is currently no method that can simultaneously measure stiffness in such diverse systems: biofilms are opaque and grow primarily in two dimensions; zebrafish embryos are transparent and develop complex 3D organs while in vivo muscle tissue sits deep under the skin and is optically inaccessible. As will be shown, OTHE allows us to study these systems using exactly the same principles of mechanical stimulation and wave analysis, providing, for the first time, maps that can be compared across species of the phylogenetic tree at different length scales and resolutions.

## Results

### Performance of OTHE

OTHE combines shear wave-based elastography, optical-flow-based wave encoding, and direct inversion of the wave equation to generate high-resolution stiffness maps. Figure 1 summarizes the assumptions in material properties, geometries, and optical properties underlying OTHE. Acquisition of shear waves in incompressible soft biological tissues by remote optical cameras required excitation units (actuators) for inducing either xy- or z-polarized waves with respect to the xy-image plane (Figure 1A). The image plane was adjusted by focussing the optical system either on the surface in opaque tissues or on the subsurface in transparent tissues. Whether in-plane waves (xy, shear horizontal - SH) or through-plane (z, Rayleigh) waves were acquired in the reflection or transmission mode depended on geometrical and optical properties as summarized in Figure 1B. Due to the small fields of view (FoVs) ranging from 2 mm to 18 mm in zebrafish embryos, adult zebrafish and biofilms, higher mechanical excitation frequencies between 800 and 2400 Hz were used, whereas *in vivo* muscle experiments were performed in the frequency range of 275 to 1000 Hz. This lower frequency range ensured the same range of wavenumbers as in the other materials, thus facilitating postprocessing with the same pipeline (Figure 1C). After synchronized serial acquisition of wave deflections, dynamic intensity correction removed areas of overexposure and underexposure from the raw intensity images. Notably, the natural surface texture and light reflectance of all studied systems provided enough speckles and features in the intensity images for displacement estimation. Wave propagation was visible to the naked eye when displayed in a cine loop after serial acquisition of optical intensity images, synchronized to the external actuator in a stroboscopic sampling scheme (see Methods). Supplementary movie 1 shows wave propagation in a homogeneous sample of ultrasound gel (in the following referred to as *phantom*) before displacement estimation with and without rigid image registration for suppression of long wavelengths which are related to compression waves. No other compression wave filter was applied downstream in our processing pipeline, highlighting the efficacy of image registration for elimination of unwanted rigid-like motion. Novel displacement estimators, optimized to time-harmonic xy- or z-waves, are detailed in the appendix. The wave signal-to-noise ratio (SNR), as calculated using the Donoho method [36], was 80 ± 5 dB, 58 ± 2 dB, and 40 ± 5 dB in phantoms, zebrafish, and biofilms, respectively, that is up to $10^5$ times higher than achieved in state-of-the-art MRE [23]. Consequently, no further noise suppression filters had to be applied before multifrequency wave inversion to convert complex-valued field components $\hat{u}_j(x,y,f)$ into maps of shear wave speed as a surrogate marker of shear modulus ($SWS^2 \sim G'$, with $G'$ denoting storage modulus).

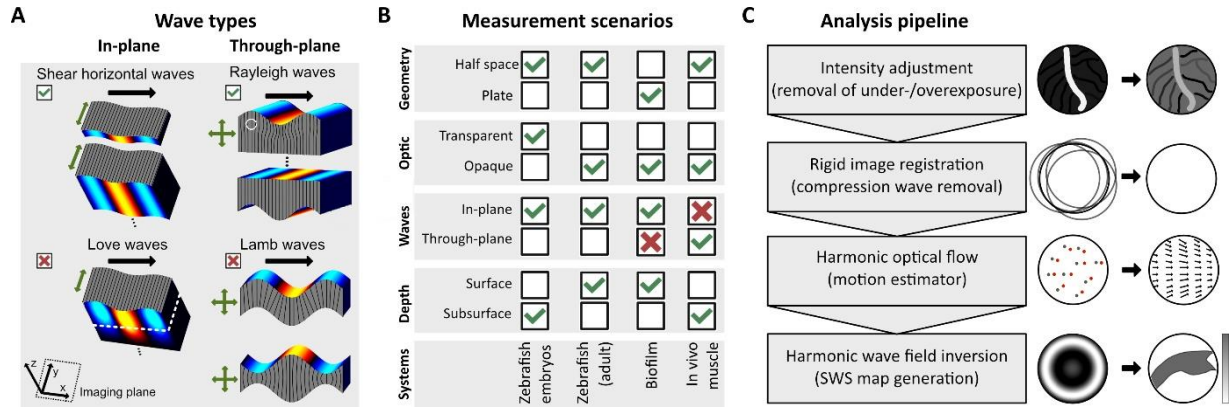

Figure 1: Principles of surface waves harnessed by OTHE, systems studied in this work, and processing pipeline. A) *Wave types propagating within half-spaces and plates including their polarization directions (green arrows) and propagation direction (black arrows) in incompressible materials. Different colors are used to represent deflection amplitudes. The orientation of the imaging plane relative to the coordinate axes is shown in the left lower corner. In the case of Rayleigh waves, the material performs elliptical motion that is governed, in incompressible tissue, by subsurface properties. Love waves propagate through layered media (dashed line demarcates the surface layer). Depending on frequency and material properties, measured SWS reflects a mixture of layer thickness and material properties of the layer and substrate. For Lamb waves, symmetric and antisymmetric modes are shown, which exhibit different SWS depending on their mode, frequency, and thickness of the plate* [37-39]. *Due to their dependence on geometry and mixed material properties, Love and Lamb waves are unfavorable scenarios in OTHE (red cross). B) Overview of different measurement scenarios to accommodate different sample characteristics. Scenarios used in this work are indicated by a green checkmark. Lamb waves are avoided because their propagation on plate thickness and deflection symmetry. Love waves are also avoided due to their strong dependence on object geometry. C) Overview of the processing pipeline used. After acquisition, images are intensity-corrected and registered to the first image to remove jitter and compression waves. Next, harmonic displacement algorithms are used to create shear wave fields, which are then inverted for generation of shear wave speed (SWS) maps. For further details, see Methods in the Appendix.*

## Phantoms

We first tested the consistency of OTHE over different excitation frequencies and spatial resolutions in a phantom. Figure 2 shows in-plane SH wave fields in a cylindrical phantom with a pixel resolution of 4 μm × 4 μm for each frequency. The apparent similarity of experimental and simulated wavefields illustrates the overall consistency of our approach (see Figure 2). Their quantitative agreement is demonstrated by line profiles and wavelength fits to estimate quantitative SWS values for all frequencies. Figure 3A also shows that OTHE values overlap with MRE values across all frequencies we used. Combining all frequencies in compound SWS maps provided consistent values in the range of the ground truth obtained by MRE (Figure 3B). Stable SWS solutions were also obtained using various lens systems with resolutions ranging from 2 μm × 2 μm to 10 μm × 10 μm, suitable for encoding SH waves with negligible inversion bias [40] in our preferred frequency range (see Methods section in the Appendix).

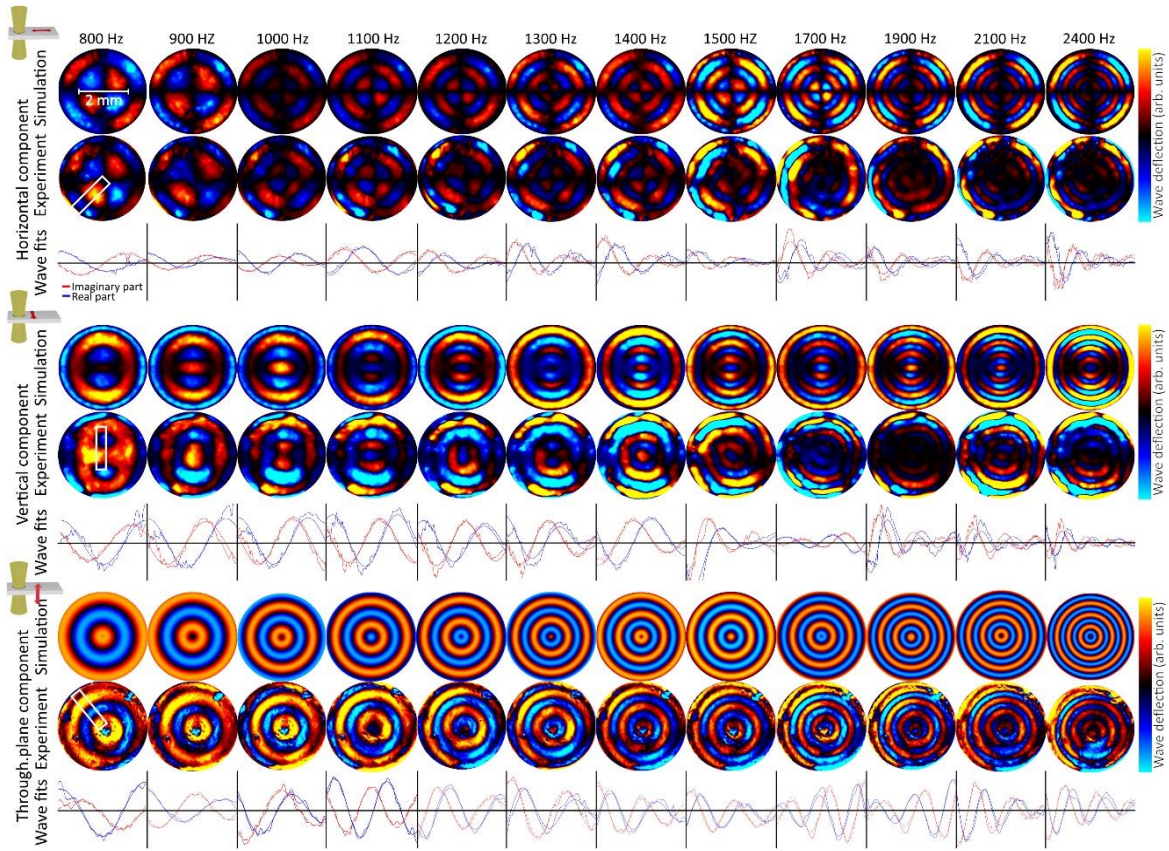

*Figure 2: Horizontal (top) and vertical (bottom) wavefield components across all frequencies measured for the ultrasound gel phantom with simulated wavefields above. A radial profile of the real (red) and imaginary parts (blue) of the waves, as used for the analyses in Figure 5, is shown below the wavefield.*

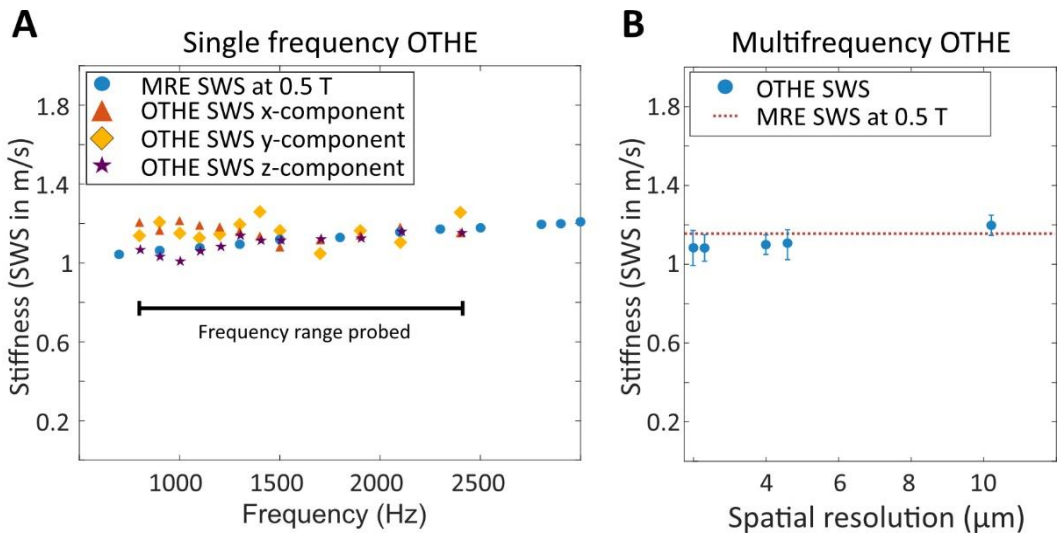

*Figure 3: A) Shear wave speed (SWS) values based on profile fits as shown in Figure 2. B) SWS values obtained by multifrequency wave inversion of OTHE images compared with 0.5 T MRE, which provided the ground truth across spatial resolutions.*

## Zebrafish

The zebrafish model [41, 42] is likely to revolutionize biomedical research due to its rapid development, large clutch size, high number of human-orthologous genes, and ease of husbandry. While zebrafish are well understood on the genetic, behavioral, and anatomic level [29, 33], biomechanical investigations of whole fish at any stage are challenging due to their small size and heterogeneity. We recently established high-resolution µMRE in the adult zebrafish [23] and used this method here as a reference for OTHE. SWS measured in the trunk muscle agreed between both modalities with 1.5 ± 0.2 m/s for OTHE and 1.5 ± 0.1 m/s for µMRE (p = 0.5), with a maximum difference of only 0.1 m/s (see Figure 4, A and B). However, µMRE provided a spatial resolution of only 60 µm, which did not allow investigation of zebrafish embryos (see Figure 4C).

Figure 4D shows camera images, wavefields of lowest and highest frequencies, and SWS maps obtained in zebrafish embryos at different ages. The wave images showed favorable characteristics for multifrequency direct inversion as they appear noise-free due to the excellent fidelity of the optical lenses in the transmission light microscope used (SNR 80 ± 5 dB). Therefore, direct inversion could be used without prior filtering, which enabled high-resolution SWS mapping. In all fish, SWS was higher in myotome than in notochord tissue (0.8 ± 0.1 m/s versus 0.7 ± 0.1 m/s, $p < 0.001$) but lower than obtained in muscles of adult zebrafish (2.9 ± 0.5 m/s). This difference could be attributable to developmental changes as suggested by the results presented in Figure 4E, where the increase in SWS in hatched fish over increasing standard length (SL) [43] is plotted. We found a linear SWS increase of 0.3300 ± 0.0003 m/s (R = 0.113, $p < 0.001$) per 1 mm standard length, which results in >1.5 m/s for 5 mm SL and explains the observed difference between freshly hatched fish and adult fish. In contrast to the development of stiffness in hatched fish, we did not observe SWS changes in either prehatch tissue (0.8 ± 0.1 m/s, $p > 0.05$) or yolk (0.7 ± 0.1 m/s, $p > 0.05$) over age or SL.

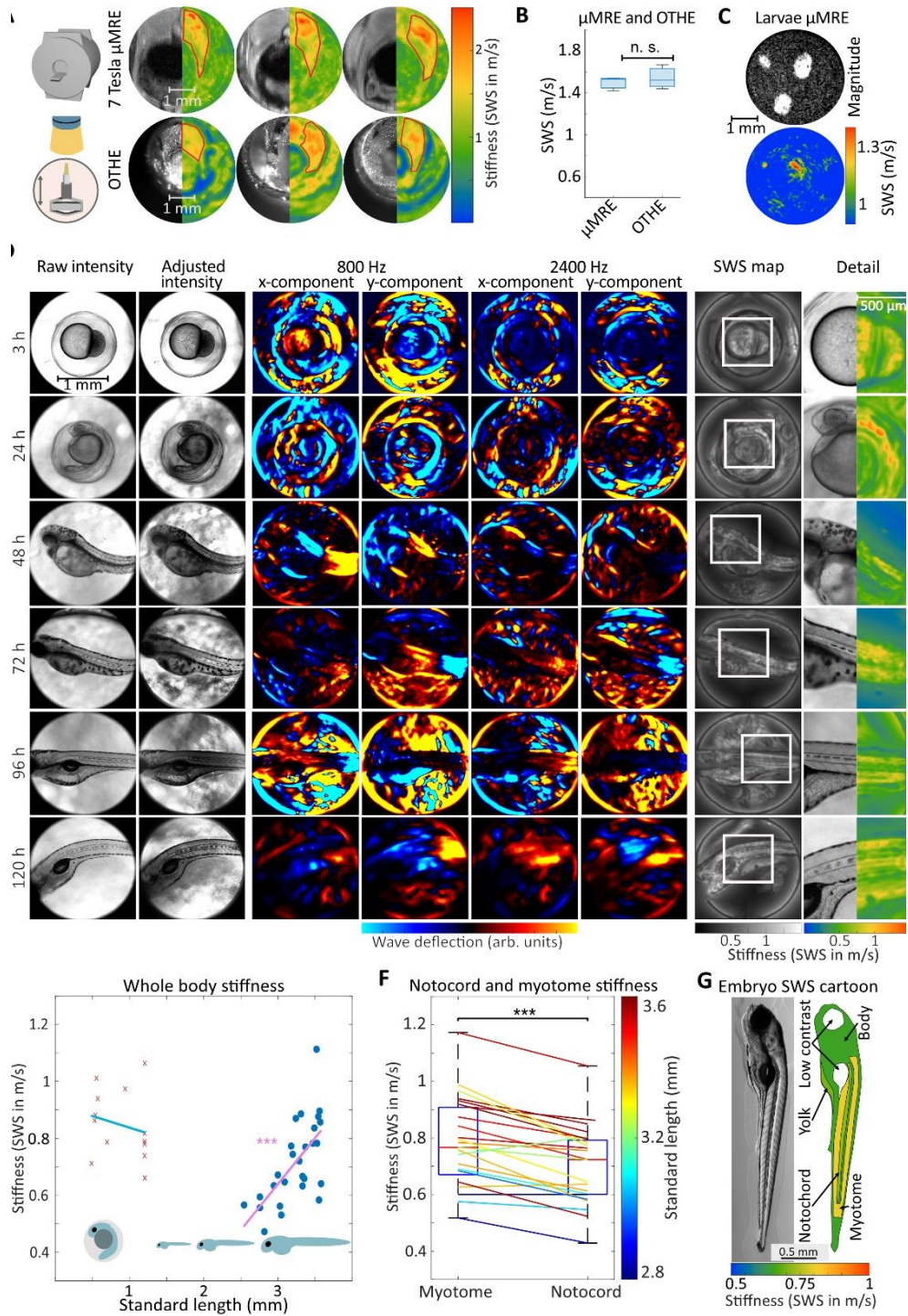

*Figure 4: OTHE in zebrafish. A) Three adult zebrafish, sectioned transversely after µMRE and examined by OTHE. The left half shows the µMRE value (top row) and the camera image (bottom row), while the right half shows the SWS maps. The muscle region is outlined in red and shows the corresponding values for both methods. B) Comparison of mean stiffness values obtained by µMRE and OTHE; differences are not significant (p > 0.05). C) µMRE images of three zebrafish embryos, with magnitude image at top and SWS map at bottom. D) From left to right: raw intensity images as captured by the high-speed camera, normalized camera images, waveforms for the lowest and highest frequency measured in both x- and y-directions, SWS map, and magnified region of interest for one sample per age group. A complete set of wave images is shown in Figure S1 A). E) Mean SWS of zebrafish embryos at different standard lengths. After hatching, stiffness of the embryos increases linearly at a rate of 0.3300 ± 0.0003 m/s·mm over the measured age range. F) Stiffness between zebrafish larval notocord and myotome after hatching is significantly different over the standard length range (p < 0.001). G) Cartoon of examined regions of interest in hatched zebrafish embryos with mean SWS values. Areas not resolvable by OTHE are shown in white. For an overview of embryo development, see Supplementary Figure 2: Development of zebrafish embryo anatomy.*

## Biofilms

Biofilms are industrially and clinically relevant [44] colonies of prokaryotes embedded in self-secreted polymeric extracellular matrix (ECM). Characteristic features of biofilms are their ability to colonize virtually any interface and their resistance to environmental stresses, including mechanical forces and effects of antimicrobial agents, which are attributable to both their material [45] and physicochemical properties [46]. Biofilms are increasingly described as analogues or sometimes even parts of human tissue [47]. The mechanical interplay of biofilms and their substrate [48] can lead to formation of a network of wrinkles that cover water-filled channels used for nutrient transport [33,48,49].

To understand this interplay between local mechanical properties and biofilm development, physiology and function, we used OTHE to map the material properties of the bacteria *Bacillus subtilis* as a biofilm-forming model. We first tested whether SH waves in the plane-strain scenario provided consistent SWS values in biofilms without being confounded by the adhesion strength to and stiffness of their substrate. Figure 5A shows biofilm SWS maps from the same sample of biofilm, both as grown on the substrate and after being remove from the substrate and measured on the bottom of a plastic dish. In both scenarios, the same vibrations were induced. Encouragingly, values were the same for the no-substrate (1.5 ± 0.3 m/s) and on-substrate case (1.5 ± 0.3 m/s), suggesting that our plane strain SH waves probed intrinsic biofilm properties and were unaffected by adhesion to the underlying support. However, substrates on which a biofilm develops can affect the biofilm's material properties [62] and growth [49]. To test whether this effect affected the stiffness measured in our experiment, we performed a series of measurements, shown in Figure 5B and D, where biofilms grown on substrates of different stiffness (1.5% and 3% agar content) were studied at three time points of 24, 48, and 72 h post seeding. The wave images presented in Figure 5 display pronounced speckles due to wave diffraction on the scale of tissue heterogeneities visible in light micrographs. Diffraction revealed motile structures and abundant slip contacts within the biofilms as a corollary of structure and composition, which might provide protection against chemical and mechanical challenges [50] (see Supplementary note 2: On spatial resolution in biphasic fluid-solid biofilms). The SWS maps shown in Figure 5D quantify the mechanical distinction between central and peripheral regions of the biofilms. The difference in stiffness between the soft central region and the stiffer peripheral region increased consistently with age for both substrates (central regions: 2.0 ± 0.3, 1.3 ± 0.3, 1.1 ± 0.3 m/s versus periphery: 1.9 ± 0.3, 1.7 ± 0.3, 1.6 ± 0.3 m/s at 24, 48, 72 h, respectively, all $p < 0.01$) with significant age-related softening ($p < 0.01$). Interestingly, while beyond 24h, there was no significant effect of substrate stiffness on the measured SWS values ($p > 0.05$), we saw an overall trend of biofilm softening with age. To cross-validate these findings we performed bulk rheology measurements of biofilms of different ages and recovered the same trend (see Supplementary note 4: Rheology measurement of biofilms).

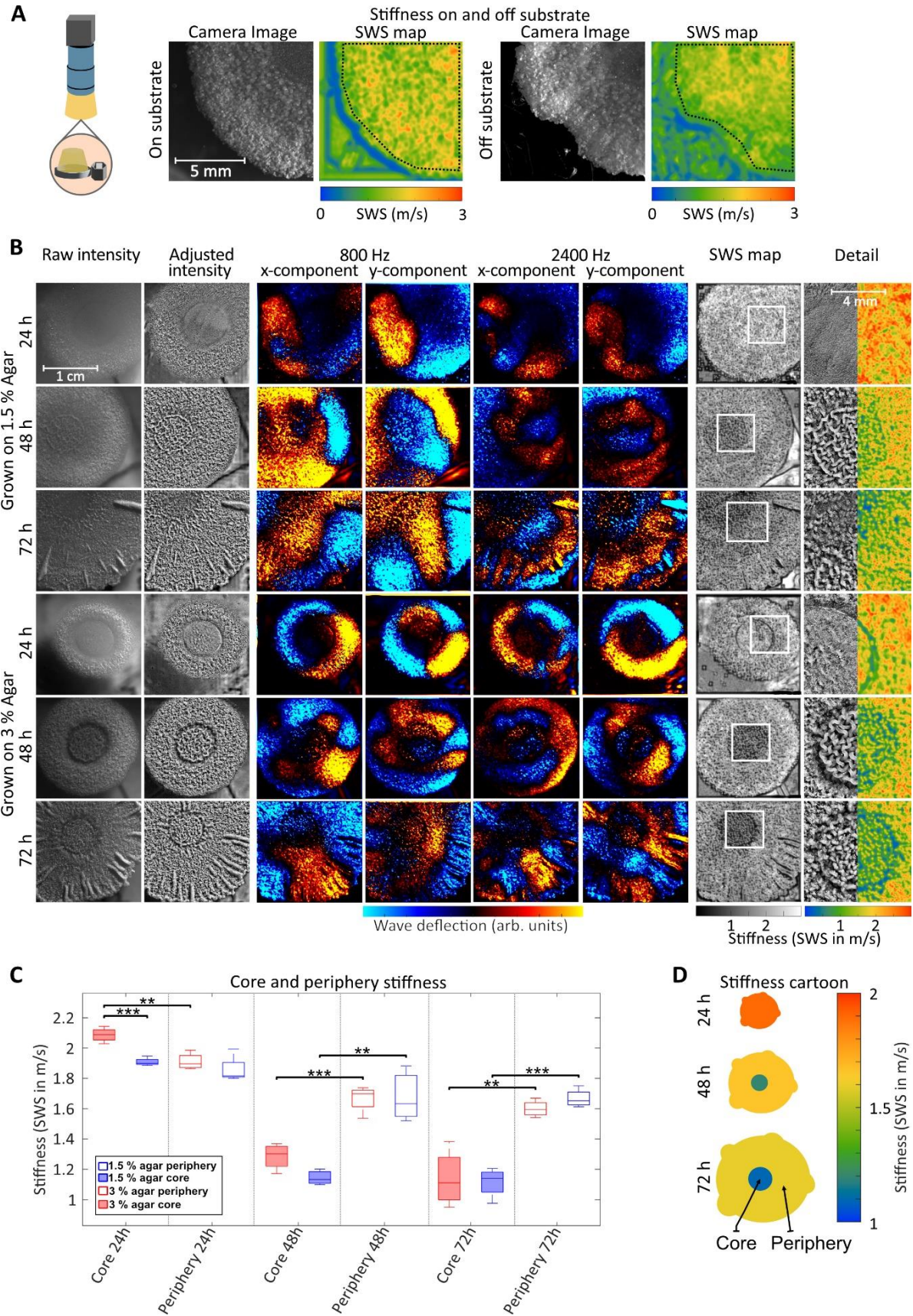

*Figure 5: A) Quadrant of a Bacillus subtilis biofilm grown on 1.5% agar for 48 hours. Left pair: biofilm measured on agar surface with SWS map on the left and camera image on the right. Right pair: the same biofilm removed from the agar and remeasured on a plastic dish. Mean SWS values for the quadrant are the same on and off the agar. B) From left to right: raw intensity images as captured by the high-speed camera, adjusted camera images, wave images for the lowest and highest frequency*

*measured in both x and y directions, SWS map and enlarged area of interest for one sample per age group and growth medium stiffness. Measurements of biofilms grown on 1.5% are shown as the top three rows, those grown on 3% agar as bottom three rows. A complete set of wave images is shown in Figure S1 B). C) Comparison of biofilm SWS in the core and periphery for biofilm grown on 1.5% agar (blue) and 3% agar (red). D) Cartoon of stiffness progression with age in both regions.*

## In vivo muscle

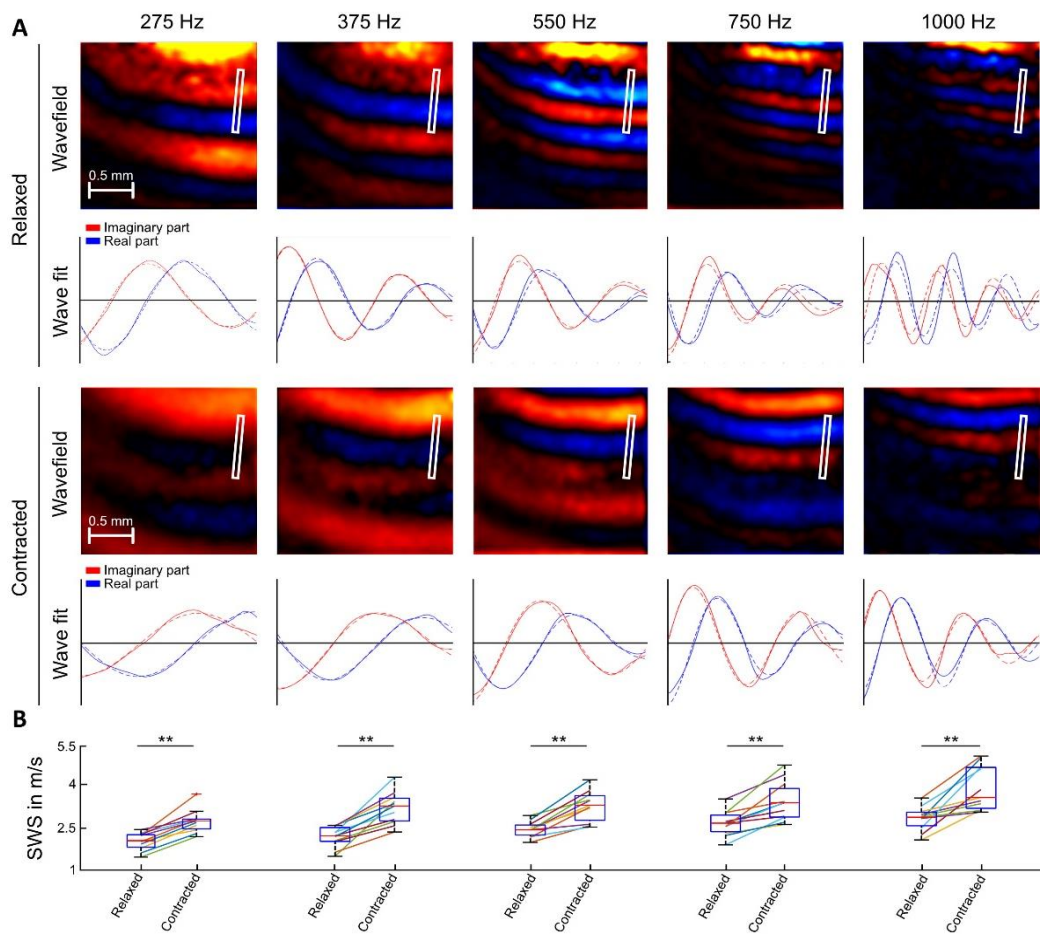

*Figure 6: In vivo OTHE in biceps brachii muscle. A) Out-of-plane wavefield components across all frequencies measured for relaxed and contracted biceps brachii. A line profile of the real and imaginary parts of the waves is shown below the wavefield, with the wave fit represented by dotted lines. B) Percent change in stiffness between relaxed and contracted muscle for all subjects and frequencies.*

The ability of skeletal muscles to change stiffness and to generate mechanical force enables the body to move. As such, muscle stiffness is intimately linked to muscle function. However, quantitative measurement of changes in muscle function due to pathology or injury remains challenging [51]. As a possible option to overcome these challenges, we tested the ability of OTHE to probe subsurface stiffness properties in vivo. This test mode was based on Rayleigh surface waves (z-deflection component), which have wavelengths in a similar range as the bulk shear waves in the underlying muscle. The encoded wave components corresponded to the slow transverse component of shear waves in transverse isotropic media (see Supplementary note 3: OTHE in anisotropic muscle tissue). Figure 6A shows the Rayleigh wavefields for each frequency for the relaxed and contracted muscle. For each frequency, the imaginary and real parts of the waves are plotted across the profile along with wave fits to determine the wavelengths and estimate stiffness. Across all frequencies, SWS values were higher in the contracted than in the relaxed muscles ($p < 0.01$) with no frequency-dependent effect size ($p = 0.4$, Figure 6B). Mean values across frequencies showed a 38% increase in stiffness with contraction (contracted: $3.4 \pm 0.8$ m/s, relaxed: $2.5 \pm 0.5$ m/s, $p < 0.01$).

## Discussion

OTHE is a remote stiffness-probing technique that provides multiscale maps of quantitative soft tissue properties for biomedical research and diagnostic applications. The method is a new approach to elastography that enables nondestructive, high-speed imaging of biomechanical properties from the tissue to the cellular resolution range. Although previously described methods of light-based elastography allow stiffness measurement at different scales, including subcellular to tissue ranges [20, 52-60], OTHE is the only cross-scale platform that can address the full phylogenetic bandwidth from bacterial cell colonies to in vivo human tissues. Based on the same time-harmonic waves that are used in MRE, OTHE offers the potential to bridge the gap between histology and in vivo medical imaging by providing multiscale SWS maps of resected tissue following in vivo MRE.

OTHE rigorously exploits a number of key innovations from seismology [61] and elastography [12], including optimized optical flow detection for time-harmonic vibration and multifrequency wave inversion for high optical SNR. MRE actuators based on piezoelectric ceramics or voice coils provide precise harmonic shear vibrations, excellent image SNR, and fast vibration in the kHz range. Although OTHE relies on tissue transparency or operates on surfaces, it is sensitive to optically inaccessible subsurface tissue when capturing Rayleigh waves. Conversely, in xy-mode, OTHE encodes SH waves that probe stiffness details both at surfaces and in the focused region of the bulk of transparent tissue. Once encoded, both types of waves impose fewer constraints on wave inversion algorithms than other elastography techniques, including MRE, for two reasons: first, rigid motion and unwanted compression waves are cancelled out using image registration. Second, due to the excellent fidelity of optical systems, OTHE SNR is orders of magnitude higher than that of MRE despite smaller pixel sizes (OTHE SNR ≈80 dB versus µMRE SNR ≈25 dB [23]versus). As a result, OTHE inversion does not require filters to suppress compression waves or noise, which is a significant advantage considering the impact of bandpass filters on resolution limits in current solutions to the time-harmonic inverse problem in elastography [40, 62].

Such a multiscale optical imaging technique, which provides µm resolution within a second of scan time, makes new experiments possible. For example, µMRE has achieved pixel edge sizes of 40 µm in 600 µm thick slices with scan times of more than 20 min for 4 frequencies [23] and cannot resolve zebrafish embryos as shown in Figure 4C. AFM achieves stiffness maps with quasi-static nano indentation of thin tissue slices and has been established as a reference modality for micro tissue mechanical examinations including those of zebrafish [63]. However, AFM maps are relatively sparse in detail because the technique relies on time-consuming point-based scanning procedures. Furthermore, AFM values are based on contact mechanics, which are difficult to translate into the range of SWS used by in vivo elastography or OTHE [14]. The maps of zebrafish embryos presented here resolve regional differences in tissue stiffness with an unprecedented level of detail. This allowed us to study stiffening in myotome and notochord in the course of embryonic development. The observed correlation of whole body stiffness over standard length is an encouraging result as it provides a linear model to interpolate developmental changes to the specific tissue mechanical state of the fish. This could also help determine at which age xenografts can be seeded in the zebrafish model at a normalized tissue stiffness independent of animal size. Consistent with our data, microdeformation analysis has previously shown that yolk sac stiffness remains relatively constant during development [64]. Brillouin microscopy has been used to assess stiffness-related parameters in zebrafish during development and regeneration after spinal cord injury. Consistent with our observation of softer notochord than surrounding tissue, it has been speculated that softness is necessary for effective locomotion [65, 66 ]. These encouraging results regarding relative stiffness changes in zebrafish tissues during disease and regeneration illustrate the importance of a quantitative stiffness mapping method for biomedical research in emerging small animal disease models.

While zebrafish larvae are too small for conventional elastography, biofilms are prohibitively thin, making OTHE the first technique for SWS mapping in such challenging systems [67]. The biomechanical properties of biofilms depend not only on the composition of the individual bacteria but also on their ECM, cell-matrix interactions as well as water distribution and content [68-70]. Therefore, the fundamental mechanisms leading to the mechanical robustness of soft tissues can be studied in biofilms using a technique such as the one presented here. Using OTHE, we here obtained the first quantitative and detailed stiffness maps of growing biofilms on different substrate layers. The observation that biofilm stiffness was unaffected by both substrate stiffness and layer contact suggests the intrinsic constitutive nature of OTHE parameters in these systems. Furthermore, the obtained stiffness contrast in our biofilm maps indicates fundamental differences between mechanical structures in central and peripheral areas (see Supplementary note 2: On spatial resolution in biphasic fluid-solid biofilms). Previous work has shown that biofilms use programmed cell death to form channels [48] that facilitate the transport of water and nutrients. Intriguingly, nutrient flux is driven by evaporation gradients formed by a larger surface area and local curvature in the center of the biofilm compared with the outer region [31]. This, combined with recent OCT studies showing stiffness differences in natural biofilms in relation to cell density [71, 72], may explain the soft core of the biofilms revealed by OTHE contrast. Our stiffness values fall into a wide range of properties reported for biofilms in the literature. Bulk rheology measurement in *B.subtilis* biofilms yielded values between 56-309 Pa (0.23-0.56 m/s SWS) [67] and 2250 Pa (1.5 m/s SWS) [73, 74] for storage modulus, which are comparable to the results we obtained by shear rheometry (see Supplementary note 4: Rheology measurements of biofilms). AFM measured a Young's modulus of 1.97 kPa (0.81 m/s SWS) in biofilms covered with ethanol (to prevent AFM tip adhesion to sticky biofilm surface), which may have induced degradation [74]. Ziege et al. reported higher values of Young's modulus of *Escherichia coli* biofilms on the order of 500 kPa (12.9 m/s SWS) without notable effects of agar substrate stiffness on biofilm stiffness in a range of agar concentrations similar to that used here [75]. In this study, higher water content was associated with lower stiffness, suggesting that an increase in ECM water due to growth causes softening. More discussion on the influence of water and water-filled channels on biofilm mechanical properties is provided in Supplementary note 2: On spatial resolution in biphasic fluid-solid biofilms. Age-related changes in water content and ECM volume fraction could also have contributed to the observed softening of the biofilms over time. OTHE is currently used in our institution to identify the specific link between the aging elements and mechanical properties in biofilms.

While biofilms and zebrafish could be studied in xy-mapping mode, OTHE operated in z-mode was sensitive to deeper tissue properties without requiring sample transparency. We chose skeletal muscle to demonstrate this sensitivity based on Rayleigh waves because muscle function is directly related to changes in stiffness, and measuring muscle stiffness in vivo remains a challenge for many current elastography methods [76-78] due to pronounced shear wave damping, anisotropy, and load-dependent stiffness variation [79]. Our preliminary study in biceps muscle was designed to provide an outlook into potential diagnostic applications. Long scan times in MRE make clinical stiffness scans at defined loading states inefficient for clinical use in this area. While ultrasound-based muscle elastography has a role in the clinic, comprehensive stiffness function tests based on combined EMG-elastography examinations have not yet been established. Surface-based wave detection methods such as OTHE could provide solutions to current challenges in diagnostic muscle imaging. Although further validation with respect to anisotropy is needed, our OTHE results fall into the range reported in the literature for the perpendicular component of relaxed muscle stiffness. For example, MRE in human biceps brachii muscle showed SWS of 2.3 ± 0.9 m/s for the perpendicular stiffness component and 5.4 ± 2.4 m/s for the parallel stiffness components at drive frequencies from 75 Hz to 118 Hz [34]. Ultrasound elastography, without consideration of anisotropy, found values ranging from 1.7 to 5.1 m/s in an unspecified range of frequencies, and identified sex, age, and elbow position as influencing factors [80].

Although this study demonstrated the basic concepts of OTHE in a rather broad range of systems, it cannot cover its full versatility in all possible scenarios. In-depth studies are needed to better understand the resolution limits of harmonic encoding and wave inversion, as well as to improve parameter consistency in human skeletal muscle by optimizing actuators, lighting conditions, and frequency ranges. Furthermore, the potential of OTHE to map multiscale mechanical parameters in overlap with in vivo MRE remains to be determined. Here, OTHE could play an important role in cancer biophysics and as a diagnostic tool based on stiffness mapping in fresh tumor biopsy specimens following in vivo MRE of the same tumor section [24]. Other potentially valuable applications, such as OTHE in zebrafish models of tissue regeneration, are also pending. Finally, beyond stiffness, time-harmonic elastography is sensitive to wave attenuation and can provide viscosity-related parameters. Although there is a large body of MRE literature on the measurement of loss modulus and damping ratio, quantification of viscosity using time-harmonic shear waves is more complex than stiffness quantification and therefore beyond the scope of this paper. The new window into soft tissue biomechanics opened by OTHE may lead to new research questions that require optimized and specialized techniques of light-based time-harmonic elastography. As such, the method presented here is a platform for further development rather than a final solution that will fit all future applications of OTHE.

In summary, OTHE is a versatile time-harmonic elastography platform that uses optical detection of surface waves or shear waves within transparent tissues. It provides solutions to several persistent problems in stiffness mapping in biomedical research over multiple length scales, including high optical resolution, ultrafast scanning, robust motion estimation of periodic tissue deflections, and multifrequency inversion of harmonic wavefields without bandpass filters. As an in-plane mapping technique, OTHE has been demonstrated to provide high-resolution stiffness maps in zebrafish embryos during development and at surfaces of growing biofilms. As a surface wave detection technique, OTHE can be used to quantify skeletal muscle function based on subsurface stiffness changes. These results illustrate the wide range of potential applications of OTHE. In addition, first stiffness reference values for the frequency range between 800 and 2400 Hz are provided for zebrafish development from egg to embryo as well as for *Bacillus subtilis* biofilms from day 1 to 3 after inoculation. Taken together, fundamental developments in shear wave generation, light microscopy, optical flow detection, and wave inversion enable OTHE to rapidly and cost-effectively map stiffness across scales in many systems relevant to biomedical research and diagnostic medical applications.

## Methods

### Optical systems

Optical intensity images were acquired using a commercial high-speed digital camera (Fastcam Mini AX100 type 540K-S, Photron, Japan) with a proprietary complementary metal-oxide semiconductor (CMOS) sensor, providing maximum frame rates from 4 kfps at full frame (1024 x 1024 pixels) to 540 kfps at 128 x 16 pixels with a minimum exposure time of 1.04 µs. The camera was controlled using the Photron FASTCAM Viewer Software (PFV4) and was set to a resolution of 1024 by 992 pixels, enabling a frame rate of 4500 fps and an ISO of 40 000. Exposure time was adjusted to the lighting conditions of the sample. Each image acquisition was independently triggered using a frequency generator (AFG 3022B, Tektronix, USA) that also supplied signal to the actuator amplifiers. The flow of data is shown in Figure 7A, and the different setups are summarized in Figure 7B.

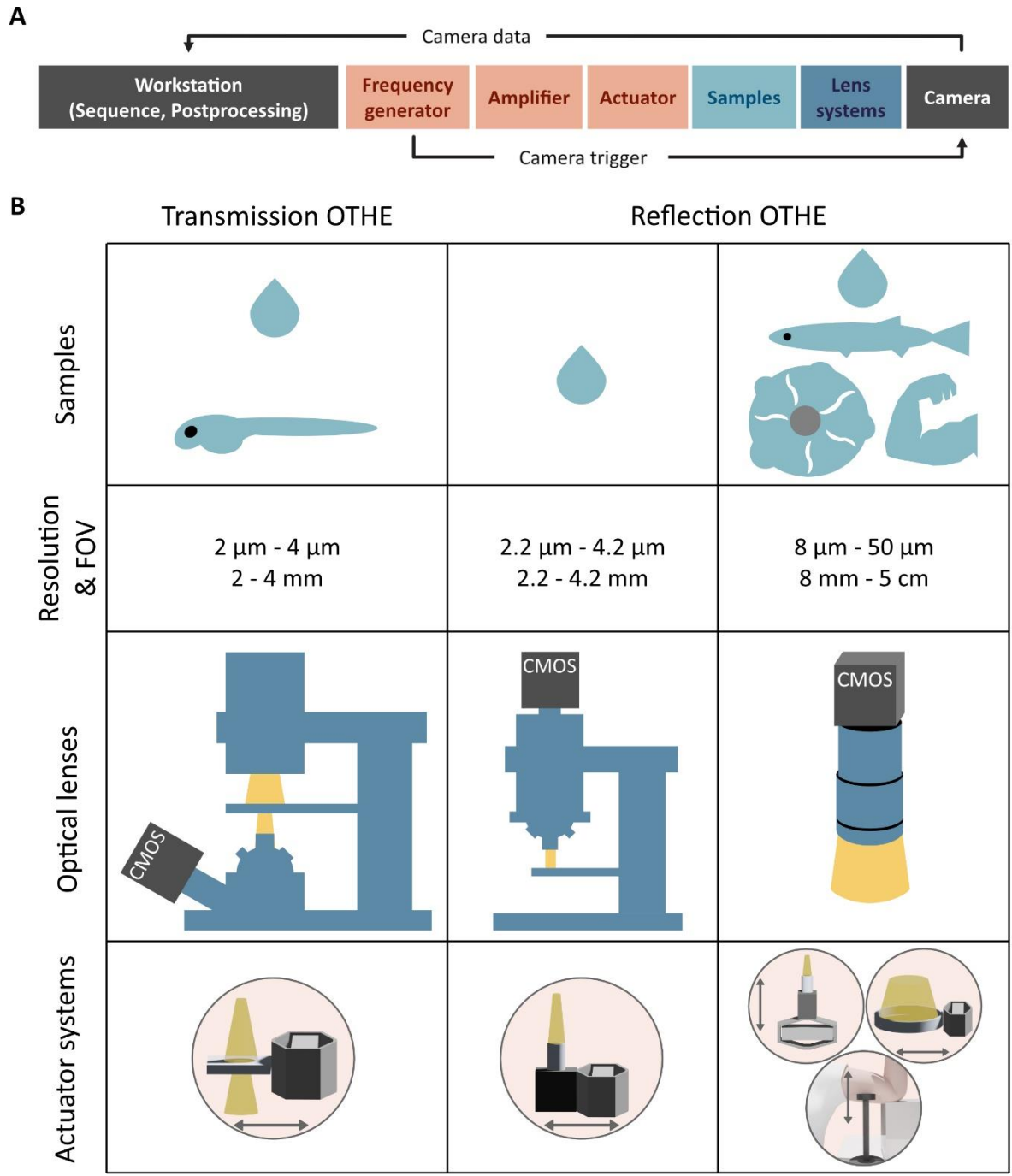

*Figure 7: A) Block diagram of setup and data flow: A set of vibrations is defined in MATLAB and exported to a frequency generator (Tektronix AFG 3022B). When the frequency generator is triggered, it sends sine waves of increasing frequency to an amplifier that powers the vibration unit. At the same time, a TTL signal synchronized to the harmonic wave triggers image acquisition. Image magnification is set by the lens system, and the acquired images are sent back to the workstation. Lens system, actuator, and amplifier can be freely exchanged to accommodate different samples and resolutions. B) Schematic overview of the experiments performed in this study. Images were acquired in transmission mode (zebrafish embryo and phantom) and reflection mode (adult fish, biofilms, and phantoms) using a microscope and macro lens, respectively. 3D drawings show the different actuator setups either based on a piezoelectric element or a voice coil (for the muscle experiments) used for each imaging modality. Arrows indicate the main vibration direction.*

As previously described, using multiple harmonic frequencies for sample excitation mitigates the effects of standing wave and wave voids in time-harmonic elastography [92, 93]. We therefore consecutively excited multiple different frequencies (800 Hz, 900 Hz, 1000 Hz, 1100 Hz, 1200 Hz, 1300 Hz, 1400 Hz, 1500 Hz, 1700 Hz, 1900 Hz, 2100 Hz, and 2400 Hz in the phantom, zebrafish, and biofilm and 275 Hz, 375 Hz, 550 Hz, 750 Hz, and 1000 Hz in human muscle). Eight images per vibration cycle were acquired at equidistant time increments over 3 cycles. To ensure steady-state time-harmonic oscillations with minimized transient deflections, we acquired no images within the first 4 vibration cycles of each frequency. The limited frame rate of 4500 fps, which is below the Nyquist frequency of some of the high-frequency oscillations, required stroboscopic sampling, as explained in Figure 7B. Here, an intended mismatch between vibration period and frame rate yielded aliased frequencies in lower spectral bins, which were precisely known due to time-harmonic particle motion. As a result, stroboscopic sampling was not limited by the maximum frame rate, which can only be increased by using larger and more expensive sensors, but by the minimum exposure time. While exposure time can easily be reduced by increasing sample illumination, the maximum frame rate is an intrinsic feature of the chosen high-speed camera and extremely expensive to increase. An overview of all experimental setups used in this study is given in supplementary table 1.

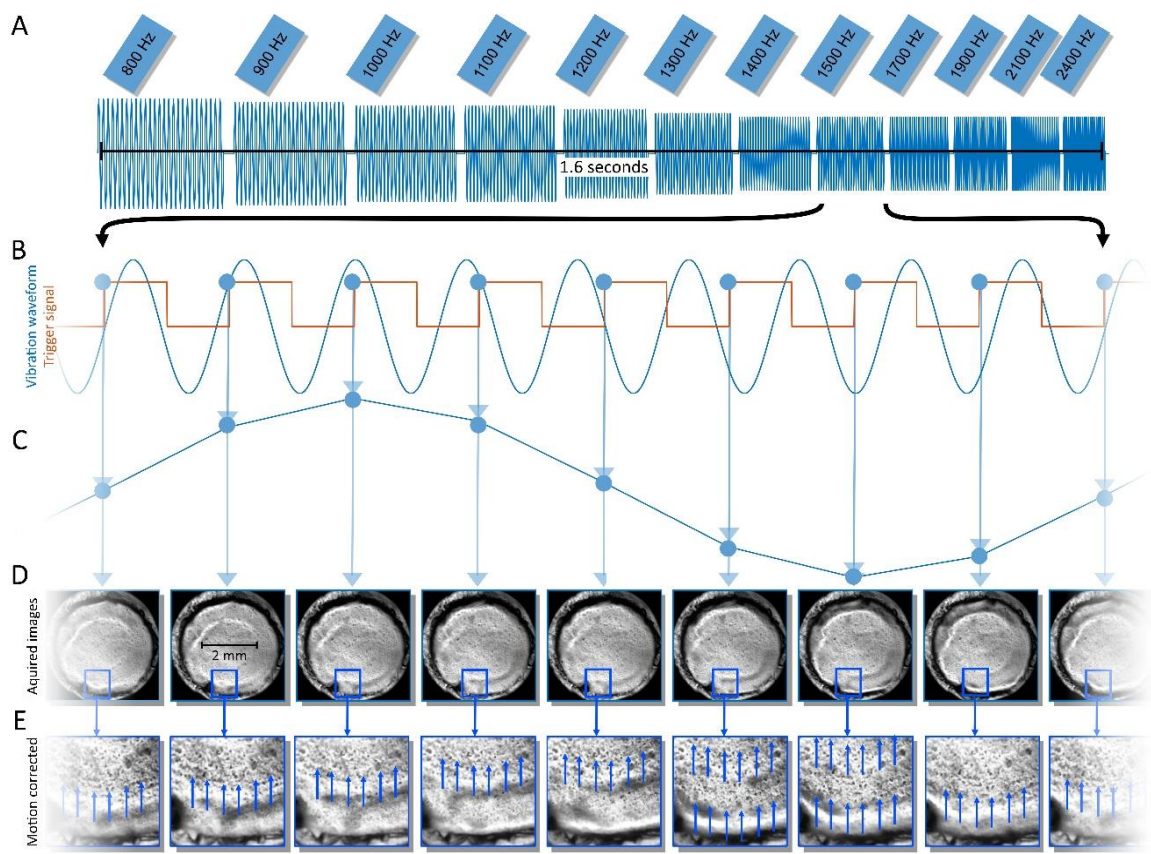

*Figure 8: Schematic overview of the image acquisition process. A) Over a period of 1.6 seconds, 12 different frequencies from 800 Hz to 2400 Hz were excited. The decrease in amplitude with increasing frequency was determined from the frequency deflection responses of the actuators. B) Enlarged view of the electrical signal sent to the mechanical excitation unit (blue) and the TTL trigger signal sent to the camera (orange). The wave is stroboscopically sampled, resulting in the reconstructed wave shown in C). D) Images acquired with the wave already visible. A magnified detail from each image is shown in E), where the propagating wavefront is clearly visible even before motion estimation.*

## Phantom

An ultrasound (US) gel (Medimex GmbH, Germany) phantom was used to validate OTHE against existing modalities. The US gel was mixed with scatterers (Siliziumcarbid F400, Mineraliengrosshandel Hausen GmbH, Österreich) to ensure good optical contrast and investigated at pixel sizes of 2 × 2 µm, 4 × 4 µm (using the Axio Observer transmission microscope from Carl Zeiss Jena, Germany), 2.3 × 2.3 µm, 4.6 × 4.6 µm (using the Axioplan reflection microscope from Carl Zeiss Jena, Germany), and 10 × 10 µm (using the LAOWA CA Dreamer macro lens, Venus Optics, China). The phantoms were exited perpendicular to the image plane by a piezo actuator (APA35XS, Cedrat Technologies, France), which was driven by a piezo actuator amplifier (PX200, PiezoDrive, Australia), and the lenses were focused on a layer below the surface. An overview of the different setups is provided in Figure 7B. Since US gel is homogenous, we were able to sidestep some of the issues presented by heterogeneous phantoms by applying global fitting methods to the measured wavefields. This was also exploited in our ground truth experiment, for which we used a previously described 0.5 Tesla tabletop MRE setup for acquisition and employing Bessel fits to extract global SWS [94].

A digital phantom was created to validate the wave images obtained. Displacement fields were simulated using 64 point sources arranged in a circle, which oscillated with the same frequencies as in the experiment, propagating waves with a shear wave speed taken from the tabletop MRE ground truth acquisition. The displacement field was then applied to a randomly populated array with the same dimensions and histogram as the acquired phantom data and sampled using the same sampling scheme as in the experiment. The images were analyzed using the standard pipeline, and the resulting wavefields are shown in Figure 2.

### Zebrafish

Zebrafish (Danio rerio) were raised and maintained according to a standard protocol at 28 °C with a 14/10 h light-dark cycle at the Pediatric Oncology/Hematology Department of Charité University Hospital, Berlin, Germany, [95]. All experiments in adult fish were performed *post mortem* as approved by the local animal ethics committee (G 0325/19, Landesamt für Gesundheit und Soziales, Berlin, Germany) and were conducted in accordance with the European Community Council Directive of November 24, 1986 (86/609/EEC). For the experiments in embryos, wild-type fish of the Tüpfel long-fin strain were used. All embryonic studies were performed post mortem with zebrafish embryos up to 5 days post fertilization and did not fall under the Protection of Animals Act.

Adult fish were euthanized immediately before examination by hypodermic shock as described in [96]. For µMRE, three adult fish (22 months after fertilization) were scanned as described previously [23]. After µMRE acquisition, we sectioned the fish transversely to make the area scanned µMRE accessible for OTHE and embedded the carcasses in the same glass tubes as employed for µMRE using US gel (4 mm in diameter). The adult fish were then mechanically excited along the body axis, through the imaging plane by the piezo actuator. We imaged the resulting SH waves at the surface using the macro lens at an exposure time of 20 µs and a pixel size of 8.3 × 8.3 µm.

In addition, a total of 42 zebrafish embryos were investigated in groups of seven at 3 h, 25 h, 48 h, 66 h, 98 h, and 120 h post fertilization (hpf). For scanning, the zebrafish embryos were embedded in US gel and placed in a 4-mm diameter bore hole covered with microscopy glass for stabilization and wave guiding. While it is possible to mount the embryos in different positions in this setup (such as those used to image the brain),, embryos were left in their natural sideways orientation in this study. The slides and embryos were excited along the image plane using the same actuator and amplifier that was used for the adult fish. Images were then acquired using the transmission microscope, exploiting the transparency of the embryos to acquire bulk SH waves with an image pixel size of 2 x 2 µm. To increase the dynamic range, images acquired at three different exposures (20 µs, 40 µs, 62.48 µs) were combined using MATLABS makehdr function. To quantify regional differences, the resulting SWS maps were then manually segmented into yolk, notochord, myochord, and full body (excluding yolk) regions of interest (ROIs).

### Biofilms

Biofilms of wild-type *Bacillus subtilis* (strain NCIB361O) were formed at a solid/air interface as follows. A liquid culture of *B. subtilis* was grown by shaking at 250 rpm for 16 h at 37 °C. A 2 µL drop of the liquid culture was placed onto plates of 1.5 % or 3 % agar-MSgg [97]. Liquid drops spread more loosely on plates containing 1.5 % agar than on 3 % agar (7.8 ± 0.5 mm droplet diameter compared to 6.1 ± 0.5 mm), possible due to slightly different drying rates of the two substrate surfaces. Plates were incubated at 30 °C for 24, 48, and 72 h prior to measurement. We scanned 12 biofilms grown on substrate containing 1.5 % agar and 12 biofilms grown on substrate containing 3 % agar. Of each group, we scanned four biofilms at 24 h, 48 h, and 72 h after inoculation. Biofilms and their supporting substrate were placed in a petri dish (2 cm diameter) that was excited horizontally using the piezo actuator and macro lens described above, with an exposure time of 33 µs and a pixel size of $18 \times 18$ µm. Since the biofilms are not transparent, we measured SH waves, which travel in the image plane. For analysis, we differentiated between the core and the growing periphery of the biofilm while excluding wrinkles growing outside the focus plane of the lens.

### In vivo subsurface muscle

The biceps brachii muscles of the right arm of 11 healthy volunteers (8 male, 3 female; age: 30 ± 9 years) were investigated after written informed consent was given. The setup of OTHE in muscle was designed to capture the slow transverse component of shear waves in transverse isotropic media (see Supplementary note 3: OTHE in anisotropic muscle tissue). Rayleigh wave motions were captured at the skin surface in a lateral position using a field of view (FOV) of $2.11 \times 2.05$ cm$^2$, a resolution of $25 \times 25$ µm$^2$, and an exposure time of 100 µs. Waves were excited outside the FOV using a voice coil (AR-50, Monacor, Germany) driven by a commercial audio amplifier (LD-Systems PA 1600-X, Adam Hall Group, Germany) in the direction transverse to the fibers (See Supplementary figure S3). Frequencies were excited at 275 Hz, 375 Hz, 550 Hz, 750 Hz, and 1000 Hz, providing similar wavelengths relative to the size of the FOV as in the other experiments conducted in this study. Two experiments were performed in each subject, once in a relaxed state of the muscle and once during maximum voluntary contraction.

### Image postprocessing

All processing steps were implemented in Matlab (R2022a, The MathWorks, USA). First, global motion correction using elastix [98] was applied to each time series of images acquired at a given oscillation frequency. This correction removed background motion and any other rigid motion component, such as compression waves, that had no notable elastic strain (Figure 8E). To remove light intensity changes, each image was normalized using contrast-limited adaptive histogram equalization included in the adapthisteq function of Matlab. Motion-corrected and normalized images were then subjected to xy-displacement estimation according to equations (11) and (12) for all experiments except muscle, which was analyzed by equation (14) for z-displacement. Multifrequency wave inversion was performed according to equation (16) using the gradient scheme described in [91] and reproduced in equation (19) with a filter width of 20 pixels. For analysis of a series of 12 frequencies with 325 images of 992 x 1024 pixels at a resolution of 10 x 10 µm, total processing time was 534 s to estimate xy-displacement and 175 s to estimate z-displacement using a quad core Intel i7-6700 with 41 GB RAM. Multifrequency wave inversion was performed pixel by pixel with negligible computational cost.

# Appendix

## Wave modes observed in OTHE

Vibrating the sample with the actuators shown in Figure 7B induces elastic deformation on the surface and inside the bulk of the sample. Figure 1A provides an overview of the different types of waves that can arise in OTHE. Given a homogeneous half-space with infinite extension in the z dimension, out-of-plane (z) deflections are generated by Rayleigh waves, whose propagation speed is within 5% accuracy of the shear wave speed (SWS) of the underlying half-space for incompressible materials [39] [61]. Thus, in first approximation, Rayleigh waves probe the shear modulus of the subsurface material. The situation is different in plates, where z-polarized waves, called Lamb waves, propagate with a speed depending on plate thickness and deflection symmetry [37]. Because of this geometry bias, Lamb waves in thin layers (such as biofilms) should be avoided. Instead, OTHE encodes in-plane shear horizontal (SH, xy-) waves. In incompressible half-space materials, SH surface waves are plane strain waves, which propagate with the same wave speed as the bulk SH waves in greater depths [37]. Therefore, SH- or xy-encoded waves are favored by OTHE as they probe the same shear modulus parameter both at the surface and in deeper tissue layers. An unfavorable scenario in OTHE that may occur in half-spaces with a surface layer is Love waves. The speed of these xy-encoded waves is a complex function of layer thickness and material properties of the layer and the half-space, making Love waves difficult to analyze without additional information on layer geometry [81]. Fortunately, Love waves only occur if SWS of the layer is smaller than that of the substrate, which is atypical on the body surface because SWS of the skin is higher than that of subcutaneous fat or muscle tissue [82]. Based on SH-encoded and Rayleigh wave modes, OTHE can measure SWS in different materials and scenarios. In the transparent zebrafish embryo, for example, focusing a microscope on a cross section through the embryo allows bulk SH wave encoding in the xy-plane using optical flow algorithms. Similarly, whenever surfaces of opaque materials are investigated, xy-encoding can be used to ensure that the measured SWS is related to the material properties of the surface layer. We therefore analyzed SH waves at surfaces in adult zebrafish and biofilms. Out-of-plane (z-) waves are favorable at layer surfaces such as the skin when probing subsurface properties based on Rayleigh wave propagation under the assumption of incompressibility.

## xy-Displacement estimation

In-plane (xy-) motion is recovered from the recorded sequence of optical intensity images $I(\mathbf{r}, t)$ based on optical flow estimation. Optical flow estimators are used to calculate motion vectors that transform one image into another [83]. OTHE assumes that the dominating motion component that is encoded in $I(\mathbf{r}, t)$, is the time-harmonic deflection field given as

$$\mathbf{u}(\mathbf{r}, t) = \text{Re}\big(\hat{\mathbf{u}}(\mathbf{r})e^{it\omega}\big) = \hat{\mathbf{u}}'(\mathbf{r})\cos(t\omega) - \hat{\mathbf{u}}''(\mathbf{r})\sin(t\omega)$$

$\omega = 2\pi f$ denotes the angular drive frequency $f$ with which the sample is vibrated. In 2D, the coordinate vector is $\mathbf{r} = (x, y)$ while $\hat{\mathbf{u}}$ denotes spatial oscillations with real and imaginary parts $\hat{\mathbf{u}}'$ and $\hat{\mathbf{u}}''$, i.e., $\hat{\mathbf{u}} = \hat{\mathbf{u}}' + i\hat{\mathbf{u}}''$. Optical flow estimates motion as $\mathbf{u} = (u_x, u_y)$ between two image frames based on Taylor series approximation of image signals. We will here introduce a method to estimate the complex-valued motion field $\hat{\mathbf{u}} = (\hat{u}_x, \hat{u}_y)$ at frequency $f$. To begin with, we adapt the most common solution to optical flow by Horn and Schunck, which minimizes the following objective function [84]:

$$E(\mathbf{u}) = E_1(\mathbf{u}) + \lambda E_2(\mathbf{u}) \qquad (2)$$

For each t, with

$$E_1(\mathbf{u}) = \int \left(\frac{\partial I}{\partial t} + (\mathbf{u} \cdot \nabla)I\right)^2 d\mathbf{r} \tag{3}$$

and

$$E_2(\mathbf{u}) = \int |\nabla \mathbf{u}|^2 d\mathbf{r} \tag{4}$$

$E_1(\mathbf{u})$ enforces constant brightness while the regularized $E_2(\mathbf{u})$ enforces spatial smoothing. Equation (2) is minimized by solution of the Euler-Lagrange equation

$$\frac{\partial I}{\partial t} \nabla I + (\nabla I \nabla I^T)\mathbf{u} = \lambda \nabla^2 \mathbf{u} \tag{5}$$

where $\lambda$ is the Lagrange multiplier. For this equation, we may consider any displacement $\mathbf{u}$, in particular, a time-harmonic motion as in (1). Inserting the ansatz $\mathbf{u}(\mathbf{r},t) = \hat{\mathbf{u}}(\mathbf{r})e^{it\omega}$ for a fixed $\omega$, and integrating over a period $T = 2\pi/\omega$, we obtain

$$\frac{1}{T}\int_{-T/2}^{T/2} \frac{\partial I}{\partial t}\nabla I\, dt + \left(\frac{1}{T}\int_{T/2}^{T/2} (\nabla I \nabla I^T)\, dt\right)\hat{\mathbf{u}} = \lambda \nabla^2 \hat{\mathbf{u}}$$

The first term on the left-hand side can be identified as coefficients of the Fourier series while the term in the middle is the mean of $\nabla I \nabla I^T$ over one period. To stabilize the solution and reduce numerical costs, we proceed with

$$m[(\nabla I)(\nabla I)^T]\hat{\mathbf{u}} + \mathcal{F}\left[\frac{\partial I}{\partial t}\nabla I\right] = \lambda \nabla^2 \hat{\mathbf{u}} \tag{6}$$

where $m$ denotes the mean over a time period and $\mathcal{F}$ is the temporal Fourier transform. The discrete Laplacian $\nabla^2$ on the right-hand side can be approximated as $\nabla^2 \hat{\mathbf{u}} \approx \overline{\hat{\mathbf{u}}} - \hat{\mathbf{u}}$ such that $\overline{\hat{\mathbf{u}}}$ denotes the weighted average in the neighborhood of $\hat{\mathbf{u}}$ at discrete positions $(k, l)$ [85]:

$$\overline{\hat{\mathbf{u}}}(k,l) = \frac{1}{N}\sum_{n=-N}^{N}\sum_{m \neq n} \hat{\mathbf{u}}(k+n, l+m) \tag{7}$$

$N$ denotes the size of the Laplacian kernel. In OTHE, $N$ is scaled with pixel size $h$ of an image by the empirically determined kernel size $N = 2^6 \mu m/h$, which ensures that the smoothing capacity of $\overline{\hat{\mathbf{u}}}(k,l)$ remains constant over a range of resolutions. Inserting equation (7) into equation (6) leads to:

$$m[(\nabla I)(\nabla I)^T]\hat{\mathbf{u}} + \mathcal{F}\left[\frac{\partial I}{\partial t}\nabla I\right] = \lambda(\overline{\hat{\mathbf{u}}} - \hat{\mathbf{u}}) \tag{8}$$

This can be solved iteratively for $\hat{\mathbf{u}}$ for each drive frequency $\omega$:

$$(m[(\nabla I)(\nabla I)^T] + \lambda \mathbb{I}) \cdot \hat{\mathbf{u}}(\omega)^{n+1} = \lambda \overline{\hat{\mathbf{u}}(\omega)}^n - \mathcal{F}\left[\frac{\partial I}{\partial t}\nabla I\right] \tag{9}$$

where $\mathbb{I}$ is the identity matrix. In each iteration step $n$, $\hat{\mathbf{u}}(\omega)$ is filtered by a median filter with the same kernel size $N$ as used for the Laplacian kernel in equation (8) [86].

### z-Displacement

Unlike xy-waves, z-deflections are 3D phenomena requiring stereo cameras, optical coherence techniques, or laser vibrometers when detected by light [60]. Alternatively, we here propose analysis of relative intensity changes due to light scattering. Light is diffracted when surfaces are displaced along the surface normal (z), resulting in apparent motion patterns corresponding to the intensity changes recorded by the optical system. Thus, after temporal Fourier transform of $I(x, y, t)$, we obtain

$$\hat{I}(x, y, \omega) \approx a_0 e^{i(\mathbf{kr}+t\omega)} \cong \hat{u}_z(x, y, \omega). \tag{14}$$

$a_0$ denotes the relative deflection amplitude along $z$ since the true physical deflection, $u_{0z}$, is obscured by image intensities. As shown below, wave inversion is based on normalized ratios of $\hat{\mathbf{u}}(\omega)$, making the absolute wave amplitude irrelevant.

## Multifrequency shear wave inversion

Following the assumption that elastic tissue properties dominate SWS, i.e., $G' > G''$, where $G'$ and $G''$ denote storage and loss modulus, respectively, of the complex shear modulus $G^* = G' + iG''$, $SWS$ can be obtained from the magnitude modulus:

$$SWS = \sqrt{\frac{|G^*|}{\rho}}. \tag{15}$$

Material density $\rho$ is assumed to be $1000 \text{ kg/m}^3$, as suggested in the MRE guidelines [87]. $|G^*|$ is obtained from direct inversion of motion field $\hat{\mathbf{u}}(\omega)$ using the Helmholtz equation, which is in magnitude representation [88]:

$$|G^*||\nabla^2 \hat{\mathbf{u}}(\omega)| = \omega^2 |\hat{\mathbf{u}}(\omega)|. \tag{16}$$

Single-frequency inversion schemes of the lossless Helmholtz equation suffer from a number of problems such as inhomogeneous illumination of the tissue due to standing wave nodes. Under ideal noise-free conditions, it would be possible to recover correct moduli even for nodes with zero displacement because Helmholtz inversion relies on spatial derivatives, i.e., analysis of the curvature of the wave. However, when noise is present, this numerical analysis is highly unstable, requiring averaging over multiple field components and $M$ excitation frequencies $\omega_m$ to stabilize the estimation of $SWS$ [89]:

$$|G^*| = \rho \frac{\sum_{j=1}^{2} \sum_{m=1}^{M} \omega_l^2 |\hat{u}_j(\omega_m)|}{\sum_{j=1}^{2} \sum_{m=1}^{M} |\nabla^2 \hat{u}_j(\omega_m)|}. \tag{17}$$

It is important to note that OTHE, unlike MRE, does not apply bandpass filters to $\hat{\mathbf{u}}(\omega_m)$ for suppression of compression waves and noise prior to inversion. Compression waves are cancelled out by rigid image registration prior to motion estimation, as shown in figure 1, while OTHE SNR is on the order of 51 to 58 dB based on the Donoho method [36, 90], which is approximately 2000 to 4000 times higher than MRE SNR. Instead of smoothing, OTHE suppresses unwanted signal by higher dimensional finite gradient schemes, as proposed by Anderssen and Hegland [91]:

$$a_{k,l} := y_{k+s_1,l} - y_{k-s_1,l} \tag{18}$$
$$b_{k,l} := y_{k,l+s_2} - y_{k,l-s_2} \tag{19}$$

$$(\nabla \hat{u}_j)_{k,l} = \frac{1}{2(2r_1+1)(2r_2+1)} \left( \frac{1}{s_1 h} \sum_{m=-r_1}^{r_1} \sum_{n=-r_2}^{r_2} a_{k+n,l+m} + \frac{1}{s_2 h} \sum_{m=-r_1}^{r_1} \sum_{n=-r_2}^{r_2} b_{k+n,l+m} \right) \tag{20}$$

$y_{k,l}$ are the pixels in the wave image at positions $(k, l)$, $h$ denotes pixel size, and $r_1, r_2, s_1, s_2$ define the range of the neighborhood in the two image directions. In this work, $s_1 = s_2 = 10$ was used.

## Acknowledgments

Support from the German Research Foundation (DFG, IS: SFB1340 Matrix-in-Vision, GRK2260 BIOQIC, CRC1540 Exploring brain mechanics and FOR5628) is gratefully acknowledged. This work is also supported by DFG Grant 504222949 (to MG, VZ and LC) in the framework of the Priority Program SPP2389 "Emergent functions of bacterial multicellularity". MG acknowledges the support of the Neubauer Family Foundation for the PhD fellowship.

# Supplementary Materials: Optical time-harmonic elastography for multiscale stiffness mapping across the phylogenetic tree

Jakob Jordan[1], Noah Jaitner[1], Tom Meyer[1], Luca Brahmè[2,3], Mnar Ghrayeb[4,5], Julia Köppke[2,3], Stefan Klemmer Chandia[1], Vasily Zaburdaev[6,7], Liraz Chai[4,5], Heiko Tzschätzsch[8], Joaquin Mura[9], Anja I.H. Hagemann[2,3], Jürgen Braun[8], Ingolf Sack[1]

1 Department of Radiology, Charité – Universitätsmedizin Berlin
2 Department of Hematology/Oncology, Charité – Universitätsmedizin Berlin
3 German Cancer Consortium (DKTK)—German Cancer Research Center (DKFZ), 69120 Heidelberg
4 The Center for Nanoscience and Nanotechnology, The Hebrew University of Jerusalem, Edmond J. Safra Campus, Jerusalem, 91901, Israel
5 Institute of Chemistry, The Hebrew University of Jerusalem, Edmond J. Safra Campus, Jerusalem, 91901, Israel
6 Department of Biology, Friedrich-Alexander Universität Erlangen-Nürnberg, 91058 Erlangen
7 Max-Planck-Zentrum für Physik und Medizin, 91058 Erlangen, Germany
8 Institute of Medical Informatics, Charité – Universitätsmedizin Berlin
9 Department of Mechanical Engineering, Universidad Técnica Federico Santa María, Santiago, Chile.

## Contents

## Supplementary note 1: Optical elastography methods

Optical methods have a long history in elastography [1]. The first optical approach for detecting compressive deformation at a micrometer scale was optical coherence elastography (OCE). This phase-sensitive method enables the detection of nanometer-scale axial displacement between two line-scans [2, 3]. With frame rates of up to 1 MHz, OCE has an impressive acquisition speed [4]. However, when strain results from static compression, the conversion of displacement into a stiffness map becomes complex [5]. Consequently, OCE has been used to track wave fronts propagated by transient acoustic impulses, a key principle in ultrasound elastography [2]. A significant advantage of transient waves is that shear wave speed can be directly estimated from acoustic phase velocity. Recently, transient waves at surfaces detected by ultrafast optical cameras have been used to explore subsurface properties across a broad range of resolutions, from single cells [6, 7] to large phantoms [8, 9]. These features highlight that optical elastography with a high-speed camera is more versatile than OCE, even though it cannot provide 3D displacement cubes of subsurface structures. Despite this limitation, surface waves polarized out-of-plane, that is Rayleigh waves, bear a strong resemblance to shear waves in the bulk of the material beneath the surface [10]. For incompressible materials, the difference in SWS between Rayleigh and bulk shear waves is only about 5%, opening a window into the assessment of subsurface properties of, for example, skeletal muscle through the skin. Moreover, capturing shear waves polarized in-plane allow stiffness mapping with the high spatial resolution provided by the optical system. However, due to SWS dispersion, stiffness is an intrinsic tissue parameter only when the dynamic stimulation frequency is known. This condition is easily fulfilled by continuously exciting time-harmonic motion and establishing a continuous flux of shear wave energy as a remote probe of tissue stiffness. Parker et al. used reverberant OCE based on single-frequency harmonic wave fields in transparent corneal tissue [11] and underneath the dura mater of the mouse brain [12] while Flé et al. detected continuous vibrations with a high-speed camera in live mouse oocytes [7]. However, these methods are single-frequency and cannot be scaled, limiting their spatial resolution for multiscale stiffness mapping.

## Supplementary Figure 1: Visualizing wave images in zebrafish and biofilm

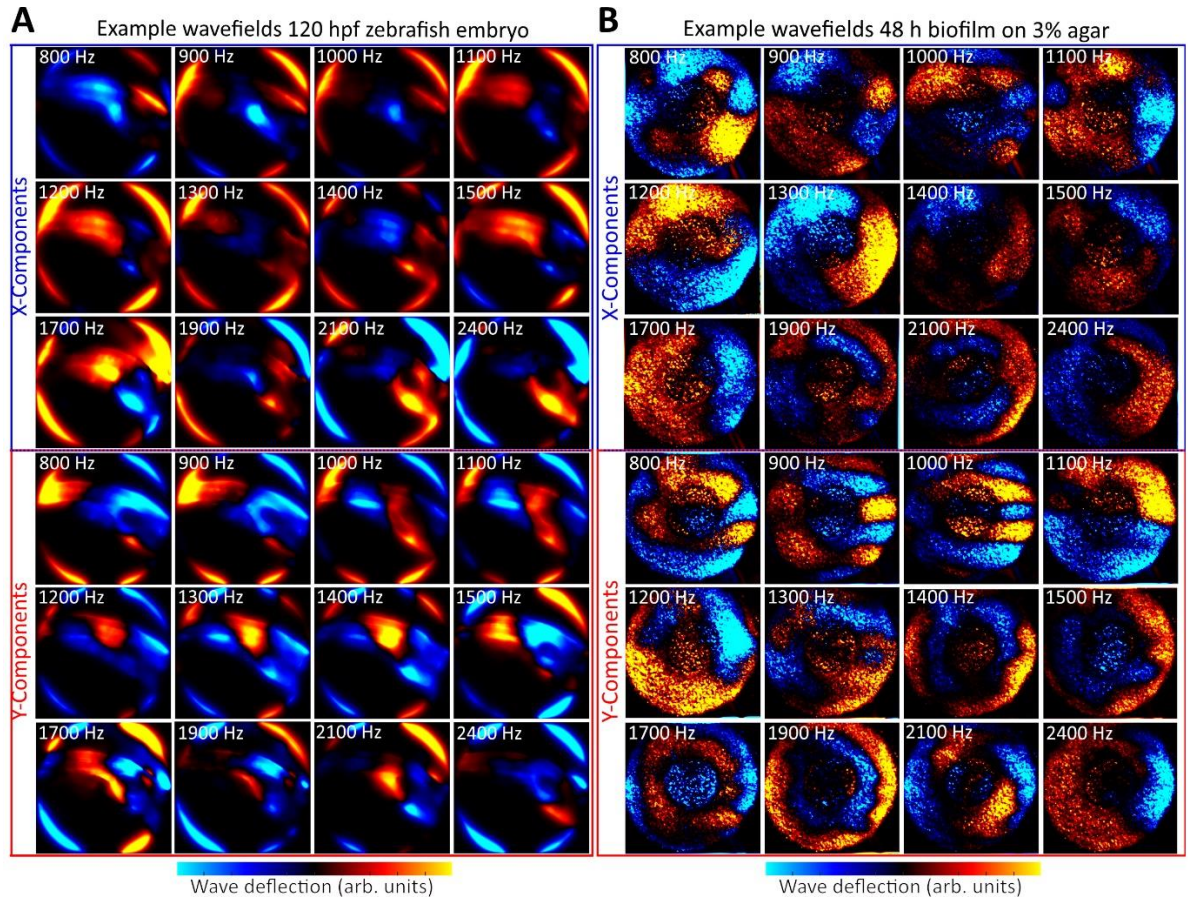

*Figure S1: Representative wave images across all measured frequencies and directions for A) one embryo at 120 hpf and B) one biofilm at 48 h of growth on a substrate containing 3% agar.*

## Supplementary Figure 2: Development of zebrafish embryo anatomy

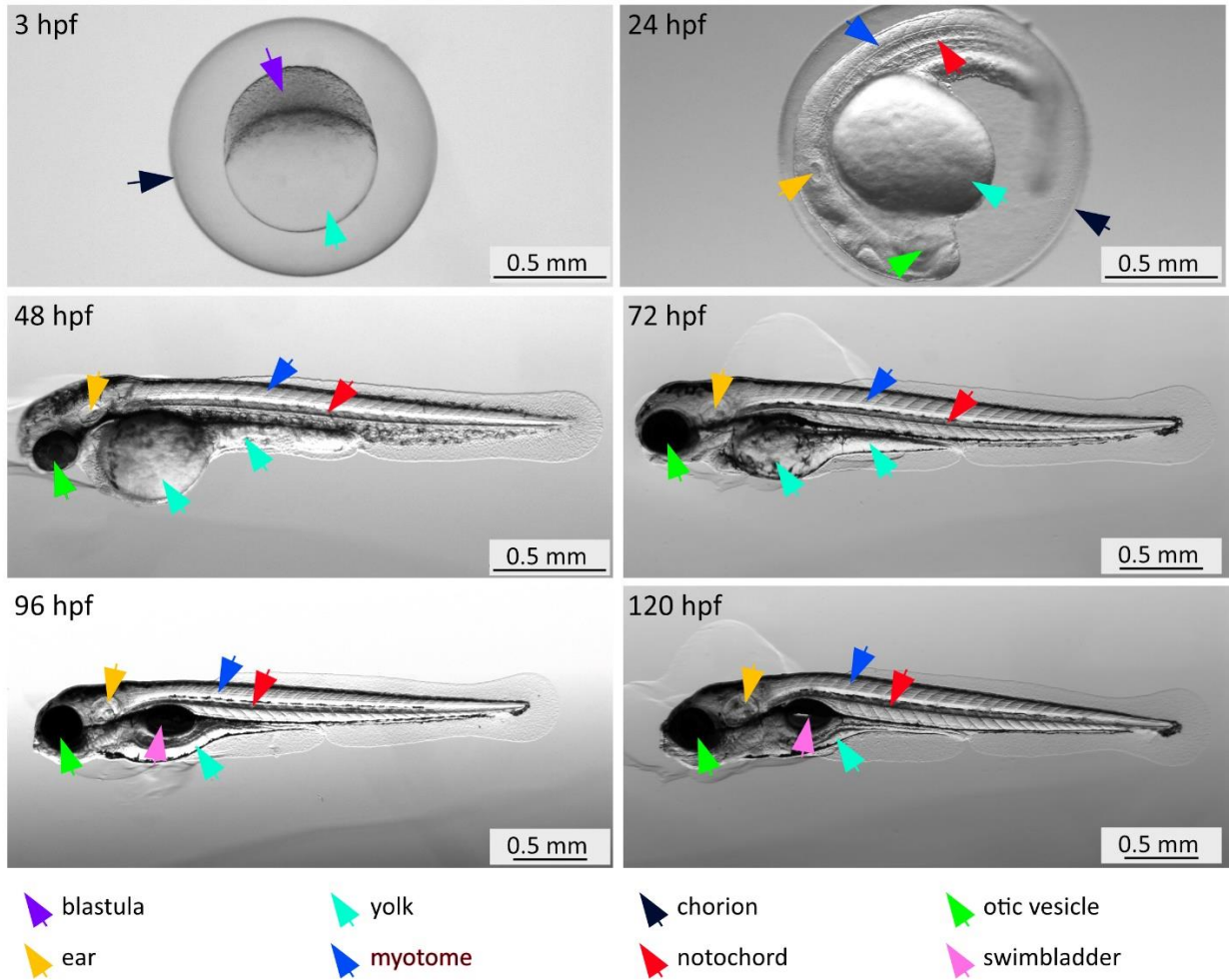

*Figure S2: Development of zebrafish embryos at between 3 hpf to 120 hpf with labeled organs.*

## Supplementary table 1: Overview of experiments with relevant parameters used in this study

| Material, tissue | Optical system | Camera pixel Resolution (μm) | Exposure time (μs) | Frequencies (Hz) | Number of time steps | Number of repetitions | Laplacian filter size (pixel) | Anderssen stencil width (pixel) | Excitation direction |
|---|---|---|---|---|---|---|---|---|---|
| Phantom | Axio Observer | 2 | 20 | 800 900 1000 | 8 | 3 | 31 | 21 | horizontal |
|  | Axioplan | 2.3 | 20 |  | 8 | 3 | 27 | 21 | horizontal |
|  | Axio Observer | 4 | 40 |  | 8 | 3 | 15 | 21 | horizontal |

|  | Axioplan | 4.6 | 40 | 1100 1200 1300 1400 1500 1700 1900 2100 2400 | 8 | 3 | 13 | 21 | horizontal |
| --- | --- | --- | --- | --- | --- | --- | --- | --- | --- |
|  | LAOWA CA Dreamer | 10 | 40 | | 8 | 3 | 7 | 21 | horizontal |
| Adult Zebrafish | LAOWA CA Dreamer | 8.3 | 20 | | 8 | 3 | 7 | 21 | vertical |
| Zebrafish Embryo | Axio Observer | 2 | 20, 40, 62.48 | | 8 | 3 | 31 | 21 | horizontal |
| Biofilm | LAOWA CA Dreamer | 18 | 33 | | 8 | 3 | 3 | 21 | horizontal |
| Muscle | LAOWA CA Dreamer | 25 | 100 | 275 375 550 750 1000 | 8 | 8 | % | 21 | vertical |

## Supplementary note 2: On spatial resolution in biphasic fluid-solid biofilms

The wrinkled surface geometry formed by biofilms is of great interest not only from a biological perspective but also from a mechanical point of view. As shown in Figure 5B and Supplementary Figure S3, the core regions exhibited abundant wrinkles of approximately 210 ± 40 μm in cross-section. In contrast, fewer but larger wrinkles of approximately 540 ± 50 μm width were visible in the periphery. These vessel-like structures in the core and periphery contain water that can move freely, as opposed to confined water compartments within cells or intercellular spaces [33]. Unconfined water within pores, wrinkles, or sulci acts as a lubricant that facilitates displacement of tissue at interfaces through shear forces [95]. As a result, a material with many slip interfaces behaves softer than the same material with welded (dry) interfaces. In shear wave elastography, slip interfaces pose a challenge because they resemble cracks in the material and cause discontinuities in the wave phase. Correct estimation of material properties near slip interfaces requires precise knowledge of the boundary conditions, which is practically infeasible. Therefore, MRE often compiles wave discontinuities into soft properties at or near the location of slip boundaries, where leaps in the phase gradient occur. Such steep wave gradients are indistinguishable from short wave numbers, which in turn encode soft tissue properties. If the crack is thin, these values are clearly incorrect. However, from a coarse-grained perspective, these values encode a boundary layer of the thickness of the point spread function of the inversion method [96]. Such a boundary layer reflects a mixture of the shear resistance of the interface and the shear stiffness of the adjacent solid compartments. Thus, stiffness at coarser resolutions reflects the sum of the microelements, including their cross-links and cracks, all of which contribute to the behavior of the material at the length scale of the measurement. Our observation of soft properties in the biofilm core appears to reflect the presence of abundant slip interfaces due to water-filled microchannels as seen in the micrographs presented in Figure S3. Reducing water content in this region would likely affect stiffness toward higher values as the lubricating effect of water decreases. Similarly, wrinkles in the periphery can be viewed as bulging and wobbly structures that are easily displaced by shear forces [96]. As a result, OTHE detects soft values in these regions. Notably, detected stiffness properties do not appear to be affected by optical properties, as suggested by the bright core area in the biofilms shown in Figure S3, while dark intensities are seen in the stiffness maps.

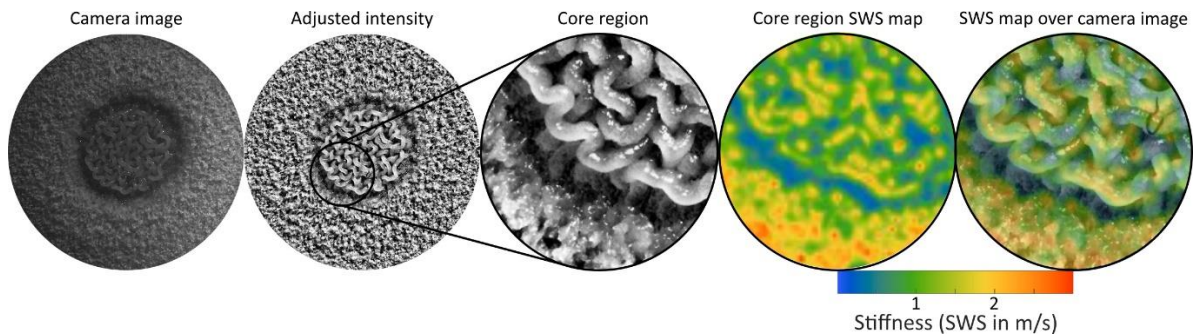

Figure S3: From left to right: acquired raw image, intensity-adjusted image, enlarged core region, and SWS map of enlarged core region. A match between stiffness and slip interfaces caused by wrinkles is visible in the SWS map.

## Supplementary note 3: OTHE in anisotropic muscle tissue

OTHE in muscle was designed to excite shear waves perpendicular to the main fiber direction. Given a transverse isotropic symmetry of the biceps brachii muscle, the propagation direction of the shear waves would also be transverse to the fibers, as shown in Figure S4. This scenario corresponds to a slow transverse wave propagation, in which shear waves probe perpendicular shear modulus $\mu_{12}$ in contrast to fast transverse waves, which are related to parallel shear modulus $\mu_{13}$.

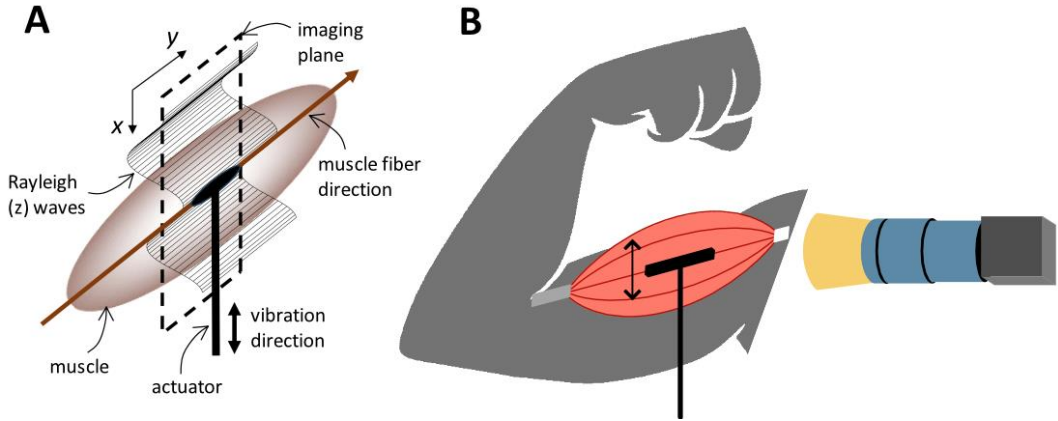

*Figure S4: A) Close-up view of biceps brachii muscle. The actuator vibrates in the x direction, perpendicular to the muscle fiber direction. The induced Rayleigh wave travels in the x direction and deflects in the z direction. B) Setup of the muscle experiment. Biceps brachii muscle shown in red. Movement of actuator indicated by black arrow.*

## Supplementary note 4: Rheology measurement of biofilms

The mechanical properties of *B.subtilis* biofilms at different ages (24h, 48h, and 72h) were measured using a rheometer (Discovery HR-2, TA Instruments). Biofilms were grown as described in the Methods section. For each acquisition, 10 biofilms were removed from their growth substrate, mixed together, and spread on the rheometer sample plate (20-mm parallel plates). Strain sweep measurements were performed at 0.01-100% strain with an angular frequency set at 10 rad/sec, followed by frequency sweep measurements performed at 0.05% strain amplitude and an angular frequency of 0.1 to 100 rad/sec. Three different sets of biofilms were scanned at each time point. The difference in absolute values between this measurement and OTHE could be attributable to homogenization of the films: complex structures like those described in Supplementary note 2 are collapsed into a uniform bulk sample, making their softness invisible to rheometer measurement.

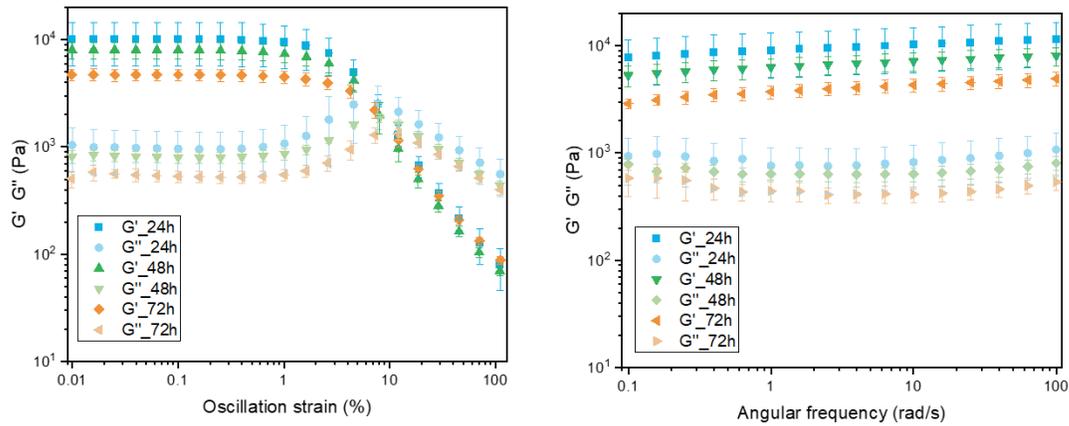

*Figure S5: Left side: G' and G'' over the range of oscillation strains probed in this study. Right side: G' and G'' for each measured angular frequency. Both measurements show a significant decrease in G' and G'' with biofilm age.*